\documentclass[prb,twocolumn,showpacs,amsmath,amssymb,floatfix]{revtex4}
\usepackage{epsfig}
\def\be{\begin{equation}}
\def\ee{\end{equation}}
\def\beqa{\begin{eqnarray}}
\def\eeqa{\end{eqnarray}}
\def\ww{\tilde{w}}

\def\nm{\nonumber}

\def\rr{\right}

\def\b{{\beta}}

\def\om{{\omega}}

\def \bp {{\bf p}}
\def \bk {{\bf k}}
\def \br {{\bf r}}
\def \bq {{\bf q}}
\def \bg {{\bf g}}
\def \bR {{\bf R}}

\begin{document}
\title{Weakly frustrated two-dimensional Heisenberg antiferromagnets: 
thermodynamic properties from a non-perturbative approach }
\author{ Leonardo Spanu$^{1,3}$ and Alberto Parola$^{2}$}

\affiliation{
 $^1$ Dipartimento di
Fisica ``A.Volta", Universit\`a di Pavia, I-27100 Pavia, Italy \\
$^2$ Dipartimento di
Fisica e Matematica, Universit\`a dell'Insubria, I-22100 Como, Italy \\
$^3$ International School for Advanced Studies, Via Beirut 4, 34013 Trieste, Italy}
\date{\today}
\begin{abstract}
We analyze the thermodynamic properties of the spin-$S$ two-dimensional 
quantum Heisenberg antiferromagnet on a square lattice with nearest and  next-nearest 
neighbor couplings in the N\'eel phase ($J_2/J_1<0.4$) employing the quantum 
hierarchical reference theory (QHRT), a non-perturbative implementation of the 
renormalization group method to quantum systems. 
We investigate the staggered susceptibility, the structure factors and the correlation 
length at finite temperature and for different values of the frustration ratio. From 
the finite temperature results, we also extrapolate ground state properties, such as spin 
stiffness and spontaneous staggered magnetization, providing an estimate of 
the extent of quantum corrections. The behavior of these quantities as a function of 
frustration may provide some hint on the breakdown of the N\'eel phase at zero temperature
for larger values of $J_2$. 
\end{abstract}
\maketitle
\section{Introduction}
Several new magnetic materials characterized by competing (i.e. {\sl frustrating}) 
interactions have been recently synthesized\cite{summa,pietro1}. 
Unfortunately, a detailed theoretical 
understanding of the ground state properties of simple models displaying both frustration and 
quantum fluctuations is still missing\cite{lul} and the interpretation of the experimental
results often relies on simple perturbative or semi-classical estimates. Needless to say,
the effects of thermal fluctuations in these systems is far from being understood either,
despite the fact that most of the experimentally accessible quantities explicitly refer to 
the temperature response. 

Among the different magnetic models with competing interactions which have been theoretically 
investigated, the so-called $J_1-J_2$ model provides a paradigm of the open problems and challenging 
questions arising in this area of theoretical magnetism \cite{lul}. 
The Hamiltonian of the square lattice Heisenberg antiferromagnet with 
nearest-neighbors ({\sl n.n}) and next-nearest neighbors interactions  ({\sl n.n.n.}) is
\begin{equation}
\label{hami}
H_J = J_1 \sum_{n.n.} \vec{S}_{{\bf i}}\cdot
\vec{S}_{{\bf j}} + J_2 \sum_{n.n.n.} \vec{S}_{{\bf i}}\cdot
\vec{S}_{{\bf j}}
\end{equation}
where $J_1,J_2>0$ and $\vec{S}$ are spin-$S$ operators. The model is characterized by a 
single parameter: $\alpha=J_2/J_1$ named {\sl frustration ratio}.
The interest in this model was triggered  by the discovery of high-T$_c$  superconductors 
and by the possible connections  between the physics of the two-dimensional Hubbard model 
and the Cu oxides materials.
It has been argued that the motion of holes in a quantum antiferromagnet gives rise to effective
frustrating couplings in the undoped Heisenberg Hamiltonian\cite{hub}, which eventually 
leads to the breakdown of N\'eel order. 
More recently, the physics of the $J_1 - J_2$ model raised a renewed interest 
due to the discovery of three vanadate compounds (Li$_2$VOSiO$_4$, Li$_2$VOGeO$_4$ and VOMoO$_4$) 
which appear to be characterized by antiferromagnetic nearest and next-nearest neighbor interactions 
between spin-$1/2$ V$^{4+}$ ions \cite{pietro1,pietro2}.

At the classical level, the zero temperature phase diagram of the $J_1-J_2$ model separates into
two regions:
for $\alpha<1/2$ the system shows  N\'eel order with magnetic wave vector $ {\bf g} = (\pi,\pi)$, 
while for $\alpha > 1/2$ the two sublattice decouple, each displaying N\'eel order, leading to
(infinitely) degenerate classical ground states. This degeneracy is lifted by quantum fluctuations 
which, at the semi-classical level, 
select collinear order with periodicity characterized by wave vectors $ (\pi,0)$ or  $(0,\pi)$. 
In recent years the search for a new disordered phase in the intermediate region of frustration
($\alpha\sim 0.5$), has been the main subject of theoretical investigation\cite{liquid1}
while the thermal properties of frustrated quantum models have attracted comparatively less 
attention. 
Finite temperature studies cannot rely on the Monte Carlo simulations, successfully 
employed in the non-frustrated case: in fact the use of these methods 
is strongly limited by the well-known sign problem, a numerical instability which occurs both in 
fermionic systems and in frustrated bosonic models.\cite{segno}.  
On the other hand, the Lanczos-based approaches do not suffer from the sign problem, but the 
limitations in the cluster size make rather difficult the calculations at 
finite temperature \cite{galliano}. 
Nevertheless thermodynamic parameters such as spin stiffness and spin velocity have been estimated 
by exact diagonalization on finite clusters followed by extrapolation to the thermodynamic limit. 
Unfortunately these small cluster results are always affected by the uncertainties in the 
finite-size scaling procedure \cite{feiguin,eina,boca}.
Among the numerical approaches, the high temperature expansion method seems to be 
the natural candidate for the analysis of thermal fluctuations\cite{hte}. 
Indeed, the physics of the collinear phase has been recently investigated by several high temperature 
expansion studies \cite{singh,misgusc}. Besides the interest related to the physics of 
the vanadate materials \cite{pietro1,pietro2,misgusc}, the study of the large 
$\alpha$-limit of the model was also motivated by the possible observation of a finite 
temperature `Ising' transition in the $J_1-J_2$ model. In fact the residual $Z_2$ symmetry 
of the Hamiltonian in the collinear phase (due to the twofold degeneracy of the ground state) 
can be broken at finite temperature\cite{chandra}, without violating the Mermin-Wagner theorem.

On the analytical side, 
Ivanov and Ivanov \cite{ivanov} investigated the low-temperature thermodynamics for $0<\alpha<1$, by 
a modified spin-wave theory (SWT). Using Schwinger-boson mean-field theory Mila {\it et al.} 
examined the spin dynamics of the model, as an effective theory for the high-T$_C$ Cu oxides \cite{mila}. 
A comprehensive study of the thermodynamics of the collinear phase, within a (semiclassical) effective 
Hamiltonian approach (PQSCHA), has been performed by Capriotti {\it et al.}, also for the 
extreme quantum case $S=1/2$ \cite{caprio}, while only a few investigations have been devoted 
to the thermodynamics of the weakly frustrated model, in the N\'eel phase. 
In the low temperature regime some hint comes from the renormalization group (RG) 
formalism applied to the non-linear sigma model (NLSM). 
Within  the  NLSM approach, the frustrating interaction is treated in mean field approximation 
and the fluctuations induced by the frustration are in fact neglected. The terms in the Hamiltonian 
acting on the variables of the same sub-lattice, such as the $J_2$ interaction, are integrated 
out in order to obtain the effective action and then the effect of the frustration reduces to 
a renormalization of the coupling constant $g$ \cite{leshouche}. As a result,
the inclusion of a next nearest neighbor interaction $J_2>0$ increases the bare coupling $g$ 
and ,following the RG picture, enhances short range fluctuations.  
Indeed, if the frustration is not strong enough  to destroy the long range order, 
the low temperature physics of the systems is believed to be the same of the non-frustrated 
case, with physical properties,such as spin stiffness and  spin velocity, renormalized by the 
inclusion of frustration. Spin-waves expansion results seems to support this picture.
However, even in the non frustrated case, the picture based on NLSM and RG approach does not adequately 
represent the experimental behavior for the quasi-2D antifferomagnet with $S>1/2$. Actually 
the agreement between theory and experiment seems to hold only for $S=1/2$ in the low and  intermediate
temperature regions. In fact, the effective Hamiltonian description is generally valid in a regime 
characterized by strong short range correlations and it is not suitable to give a detailed 
account of the physics in a range of temperature where local order weakens or even disappears. 

In order to describe the interplay between frustration, temperature and quantum fluctuations, 
we follow here a different method which aims to reconcile the RG approach with a microscopic
description of the model. This theory, named Quantum Hierarchical Reference Theory (QHRT)\cite{ap} 
does not rely on a coarse graining procedure leading to some effective action but instead is a 
way to implement the momentum space renormalization method directly to the physical Hamiltonian. 
Starting from the mean field approximation, we recursively include the fluctuations over all 
length scales by means of an evolution equation, which describes 
how the free energy of the system changes when fluctuations over growing wavelengths are taken into account.
The original method developed by Wilson has been largely employed for the investigation
of the properties of quantum systems, but, contrary to the weak coupling RG method
applied to effective field theories, our approach does not require any mapping on coarse grained
actions and applies to all temperature regimes.
The resulting evolution equation is formally exact but it is not written in a closed form: 
An ansatz for the momentum and frequency dependence of the dynamical structure factor
is needed in order to put the equation in a closed and solvable form.  
The choice of a parameterization for the spin-spin correlation function is the 
only approximation present in this approach. 

The paper is organized as follows. In Section $II$ we derive the evolution equation for 
the $J_1-J_2$ model. In Sec. $III$ we investigate the effects of the frustration 
on the spontaneous magnetization and other ground state properties, like spin stiffness.
In Sec. $IV$ we study the finite temperature properties: 
susceptibility, correlation length and specific heat.  
In the last Section we comment on the implications of our results. 

\section{QHRT EQUATION FOR THE $J_1-J_2$ MODEL}

We briefly summarize the basic steps followed for the derivation of the QHRT evolution equation. 
A detailed description of the method can be found in Ref. \cite{ap}.
Being interested in a parameter region where antiferromagnetic correlations dominate,
it is convenient to include in the Hamiltonian a {\sl staggered} external magnetic field
which directly couples to the N\'eel order parameter.
The Hamiltonian is then written in terms of a reference part $H_0$, which in our case 
just coincides with the external field, and the interaction 
term $H_J$, chosen as the Hamiltonian of the $J_1-J_2$ model
\begin{equation}
H =H_0 + H_J = -h \sum_{\bR} e^{i\bg \cdot \bR} S_{\bR}^{z} + H_J
\end{equation}
where $\bg$ is the antiferromagnetic wavevector and $H_J$ is defined in (\ref{hami}). 
With our choice of $H_0$, the properties of the reference system are known.
The method applies to all interactions $H_J$, assumed bilinear in some 
operator $\rho(r)$ ,i.e. $H_J=\frac{1}{2}\int dx dy \rho(x)w(x-y)\rho(y)$; 
in the present case $\rho(r)$ represents local spin variables and
the interaction in Fourier space is $\ww(\bk)=2 J_1[\cos k_x + \cos k_y] + 4 J_2 \cos k_x \cos k_y$.
As a first step we write a formal perturbative expansion of the partition function of 
the model $Z=Tr\exp(-\beta H)$ in powers of the interaction $\ww$ \cite{fetter}.  
Here we do not make use of any specific property of the reference system correlation functions
but instead the diagrammatic perturbative expansion is carried out in full generality.
Remarkably, the resulting perturbation series is formally equivalent
to that of a classical system in $(2+1)$ dimensions, with the periodic ``time" variable
now belonging to the interval $(0,\beta)$ and the interactions described by the 
classical two body potential $w_c(\br,t)=w(\br)\delta(t)/\beta$. Mapping on classical systems
are frequently employed to describe quantum models at phase transitions, where 
long wavelength fluctuations always dominate.
However, we stress here that our approach does not rely on any continuum limit
and preserves the microscopic character of the Hamiltonian: actually, 
the formal expansion of the partition function leads to a quantum-to-classical 
correspondence which is exact, valid in the whole temperature range and not limited to 
the critical regime. 

The thermodynamics of the corresponding classical system can be studied within the Hierarchical 
Reference Theory (HRT) formalism\cite{ar}, successfully employed in the framework of 
classical statistical physics. HRT has been developed as a way to implement the basic ideas of
the momentum space RG directly to microscopic classical Hamiltonians.
The main goal of HRT is to describe how the inclusion of fluctuations over different length 
scales, affects the free energy of the model. For this purpose we define a sequence of 
auxiliary systems (named $Q-systems$) characterized by a cut-off dependent 
potential $\ww(k)_Q$ which coincides with the physically relevant one $\ww(k)$ for $k>Q$ 
and vanishes elsewhere \cite{note}. 
Introducing a momentum cut-off in the interaction inhibits the antiferromagnetic
fluctuations at long wavelengths, preempting the occurrence of phase transitions. When the
cut-off $Q$ is reduced, the properties of the system continuously evolve between a mean field
description, where fluctuations are absent, and those of 
the fully interacting system, where the fluctuations over all length scales are taken into 
account. Following this route, we obtain an exact set of differential equations expressing how  
the free energy $a^Q$ and the correlation functions change when the cut-off $Q$ 
is varied. In particular, the evolution of the Helmholtz free energy is given by
\begin{widetext}
\beqa  
\frac{d \, a^Q}{dQ}&=&\frac{1 }{2\b} 
\sum_{\om_n}
\int_{\Sigma_Q} \frac{d^2 k}{(2\pi)^2}\,\,
\,\Big \{
 \ln (1-F^{zz} _Q(\bk,\om_n)\ww(\bk))+
\label{free} \\
&& \ln \left [ (1+F^{xx} _Q(\bk,\om_n)\ww(\bk))
(1+F^{xx} _Q(\bk+\bg, \om_n)\ww(\bk+\bg))+ F^{xy} _Q(\bk,\om_n)F^{xy} _Q(\bk+\bg,\om_n)  
\ww(\bk)\ww(\bk+\bg) \rr] \Big \}\nm
\eeqa
\end{widetext}
The summation is over the Matsubara frequencies $\omega_n = 2\pi n/\beta$
and the integration is restricted to the one dimensional domain $\Sigma_Q$ defined by 
\be
Q = \sqrt{1 -\frac{1}{2}\left [\cos(k_x)^2+\cos(k_y)^2\right ]} 
\label{sigma}
\ee
with $Q\in [0,1]$. Physically, the right hand side of this differential equation provides the
contribution to the free energy due to fluctuations characterized by
a wave vector belonging to the domain  $\Sigma_Q$, which in fact spans the whole 
Brillouin zone when the cut-off parameter $Q$ varies from $1$ to $0$.
The choice of the sequence of surfaces is in principle arbitrary but 
definition (\ref{sigma}) proves quite convenient because it respects the 
symmetries of the lattice, guarantees the
stability of the partial differential equation (\ref{free}) and leaves the 
critical fluctuations, of wave vector $\bg$, at the end of the integration. 

The key quantities appearing in Eq. (\ref{free}) are 
the spin-spin dynamical correlation functions of the $Q$-system
$F^{\alpha \beta}_Q(\bk,\om_n)=<S^{\alpha}(\bk,\om_n)S^{\beta}(-\bk,-\om_n)>$.
When $Q=1$ no fluctuations have been included yet and 
the spin-spin correlation functions reduce to their expressions in 
mean-field  approximation, obtained by keeping only the zero-loop contributions in the diagrammatic
expansion:
\beqa
&& F^{xx}_{Q=1} (\bk,\omega) =\frac{\mu_\perp+\ww(\bk+\bg)}{
m^{-2}\omega^2+ (\mu_\perp +\ww(\bk+\bg))(\mu_\perp + \ww(\bk))}   \nm  \\
&&  \nm \\
&& F^{xy}_{Q=1}(\bk,\omega)=\frac{m^{-1}\omega}{
m^{-2}\omega^2+ (\mu_\perp +\ww(\bk+\bg))(\mu_\perp + \ww(\bk))}   \nm \\
&& \nm \\
&& F^{zz}_{Q=1}(\bk,\omega)=\frac{\delta_{\omega,0}}{\mu_{||}+\ww(\bk)}
\label{correl}
\eeqa
where $\mu_\perp$ and $\mu_{||}$ are known functions of temperature $T$ and 
staggered magnetization $m$. For $S=1/2$:  $\mu_\perp=2\,T\,m^{-1}\tanh^{-1}(2m)$ and 
$\mu_{||}=4\,T\,(1-4m^2)^{-1}$. The mean field form (\ref{correl}) of the correlation functions 
corresponds, after Wick rotation, to the single-mode approximation for the dynamic structure factor
\begin{equation}
\label{dina}
{\rm Im} S_{xx}(\bk,\omega)=\frac{\omega}{1-e^{-\beta \omega}}\frac{\delta(\omega-\epsilon_{\bk})+
\delta(\omega +\epsilon_{\bk})}{\mu_\perp + \ww(\bk)}
\end{equation}
where the quasi-particle dispersion $\epsilon_{\bk}$ is given by
\begin{equation} 
\label{dispersione}
\epsilon_{\bk}=m \sqrt{(\ww(\bk)+\mu_{\perp})(\ww(\bk + \bg)+\mu_\perp)}. 
\end{equation}
As pointed out in ~\cite{ap}, a better parameterization of the spin-spin dynamical correlation function (\ref{correl}) is in principle possible, allowing more complex dispersion relations. Here, for simplicity and in order to verify the reliability of the method for frustrated systems, we consider the simpler approximation reproducing the lowest order in SWT.
Linearization of Eq. (\ref{dispersione}) around $\bk=0$ shows that in mean field approximation, 
the spin velocity $c_s$ is proportional to the staggered magnetization $m$: 
$c_s=m\,J_1\,\sqrt{8(1-2\alpha)}$, which correctly reproduces only the zeroth order SWT. 

For a generic value of the cut-off $Q\ne 1$, $F^{\alpha\beta}_{Q}(\bk,\omega)$ 
are not known because they already include the contribution of short range correlations.
As a consequence, equation (\ref{free}) cannot be solved by straightforward integration
and additional information on the form of the dynamical correlation functions is required. 
One possibility is to write an infinite set of exact differential equations also for the 
many particle correlation functions, thereby generating
a full hierarchy of evolution equations: the QHRT. However, it is more convenient to limit our attention
to the first equation of such a hierarchy, i.e. to Eq. (\ref{free}), attempting a
closure by imposing a suitable {\sl approximate} parametrization of the spin-spin correlation function. 
The simplest choice, which we will adopt in this work, is to retain the same analytical 
structure of the mean-field approximation (\ref{correl}) keeping $\mu_\perp$ 
and $\mu_{||}$ as free, cut-off dependent 
parameters. Their initial value at $Q=1$ is set by the mean field expressions, while at $Q\ne 1$ they
are obtained by use of the exact sum rules which relate the dynamical correlation functions
to derivatives of the free energy $a^Q$:
\be
\label{close1}
\left[\frac{\partial a^Q}{\partial m}\right]^{-1} m=F^{xx}_Q(\bg,\omega=0)
= [\mu_\perp^Q - 4(J_1 - J_2)]^{-1}
\ee
\be
\label{close2}
\left[\frac{\partial^2 a^Q}{\partial m^2}\right]^{-1} =F^{zz}_Q(\bg,\omega=0)
=[\mu_{||}^Q - 4(J_1 - J_2)]^{-1}
\ee
Within such an approximate closure, the inclusion of fluctuations does not change the 
analytical structure of $F^{\alpha \beta}_Q$: fluctuations lead to a renormalization of the 
parameters  $\mu_\perp^Q$ and $\mu_{||}^Q$ but
do not allow for the creation of an incoherent contribution to the dynamical structure factor
(\ref{dina}). 
By use of the representation (\ref{correl}) in the exact evolution equation 
(\ref{free}) and summing over the Matsubara frequencies, we obtain an approximate, closed evolution 
equation for the free energy $a^Q$ as a function of the cut-off $Q$ and of the 
staggered magnetization $m$: 
\begin{widetext}
\begin{equation}
\frac{d \, a^Q}{dQ}=\frac{1 }{2\beta}
\int_{\Sigma_Q}\frac{d^2 p}{(2\pi)^2}\, 
4\,\ln \left [\frac{\sinh\left (\frac{1}{2}\beta m\mu_\perp\right )}{
\sinh\left (\frac{1}{2}\beta m\sqrt{(\mu_\perp+ \ww(\bp))(\mu_\perp+ \ww(\bp+\bg))}\right )}\right ]
+ \ln\left [\frac{\mu_{||}^2}{(\mu_{||}+\ww(\bp))(\mu_{||}+\ww(\bp+\bg))} \right].
\label{evoluzione}
\end{equation}
\end{widetext}
Equation (\ref{evoluzione}), together with the thermodynamic
consistency relations (\ref{close1},\ref{close2}) and the mean field initial condition at $Q=1$, 
has been solved numerically for different values of the temperature $T$ and spin $S$.
The $Q$-dependence of the parameters $\mu_\perp$ and $\mu_{||}$ is understood in Eq. (\ref{evoluzione}).
We recall that $Q$ is just 
a control parameter which allows for a gradual introduction of magnetic fluctuations:
All physical quantities we are interested in refer to the $Q=0$ limit. However, the
numerical determination  of the free energy at $Q=0$ requires the integration of 
the parabolic partial differential equation in the whole interval $0<Q<1$.
All the numerical results shown in the following sections will refer to $Q=0$.
In order to simplify the notation, the label $Q$ will be dropped when we refer to $Q=0$.
We observe that the form of equation (\ref{evoluzione}) is unaffected by $S$ and the 
spin value just enters through the initial condition at $Q=1$.
Equation (\ref{evoluzione}) has been written in quasi-linear form and solved by a 
fully implicit finite difference scheme. We stress that Eq. (\ref{evoluzione}) 
has been solved only at non-zero temperature and we did not attempt a direct analysis
of the limiting equation as $T\to 0$.

\section{ZERO TEMPERATURE LIMIT}

\begin{figure}
\vspace{2mm}
\begin{center}
\epsfig{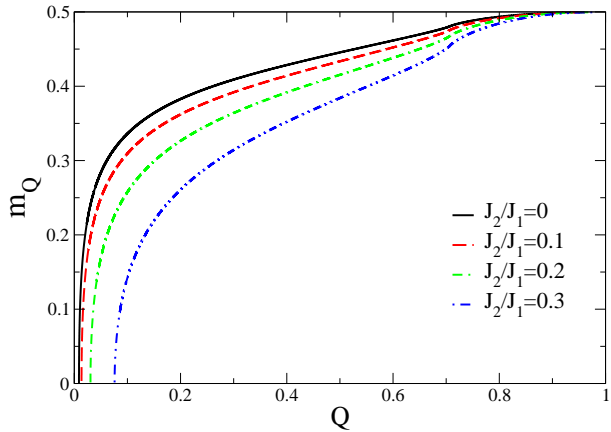}
\caption{$Q$-evolution of the zero field staggered magnetization for $S=1/2$ and different 
values of the frustration, at $T=0.1 J_1$.  
\label{evolQ}
}
\end{center}
\end{figure}

We first investigate which information on the ground state properties 
of the frustrated antiferromagnet may be extracted form the QHRT formalism. 
Let us consider the behavior of the zero field staggered magnetization when 
fluctuations are included.
By differentiating Eq.(\ref{evoluzione}) with respect to the 
magnetization it is possible to obtain an evolution equation for the staggered 
field $h_{Q}(m)$, defined as the derivative of the free energy $a^Q$ 
with respect to the staggered magnetization, at fixed $m$. This quantity implicitly describes the 
dependence of the staggered magnetization $m_Q$ on the cut-off $Q$ at fixed magnetic field $h$
through the identity $h_Q(m_Q)=h$.
\begin{table} [b]
\begin{tabular}{|c|c|c|c|c|c|c|}\hline
$\alpha=J_2/J_1$     &$0$ &$0.1$ &$0.2$ &$0.3$ &$0.35$ &$0.4$  \\
\hline
$m_0$ ($\textstyle{S=1/2}$) &$0.35$ &$0.34$ &$0.30$ &$0.25$ &$0.18$ &$0.05$  \\
\hline
$m_0$ ($\textstyle{S=1}$) &$0.84$ &$0.81$ &$0.75$ &$0.64$ &$0.52$ &$0.34$  \\
\hline
\end{tabular}
\caption{Ground state staggered magnetization $m_0$ for $S=1/2$ (top panel) and $S=1$ (bottom panel).
\label{magtab}
}
\end{table}
Note that in mean field approximation the model displays a finite temperature 
critical point between a high temperature paramagnet and an antiferromagnetic ground state. 
As a consequence, the initial condition of our QHRT equation shows non vanishing spontaneous 
magnetization $m_{Q=1}\ne 0$ for $h=0$. According to the Mermin-Wagner theorem, thermal fluctuations
depress, and eventually kill, long range N\'eel order and, as we know, QHRT does conform to this scenario
\cite{ap}. Therefore, we expect that, at low temperature, the $Q$ evolution of the 
staggered magnetization in zero field is characterized by a smooth initial behavior 
which extrapolates to the $T=0$ order parameter, followed by a sudden drop close to $Q=0$
due to the action of long wavelength fluctuations. This expected qualitative behavior can be 
checked on the numerical solution shown in Fig. \ref{evolQ}. Note that the QHRT equations 
preserve the convexity of the free energy in the physical limit $Q=0$: This implies that at
all temperatures $T\ne 0$, the zero-field inverse susceptibility $\chi^{-1}$ 
must be strictly positive at $Q=0$ and then it has to cross $\chi_Q^{-1}=0$ at a non vanishing
cut-off $Q^* \ne 0$. As a consequence, also the spontaneous magnetization, present at mean 
field level even at $T\ne 0$, vanishes at $Q^* \ne 0$ as shown in Fig. \ref{evolQ}. 
The main effect of frustration on the evolution of the spontaneous magnetization is in 
fact to move this characteristic cut-off toward larger values
due to the expected decrease of the correlation length. 

\begin{figure}
\vspace{2mm}
\begin{center}
\epsfig{figure=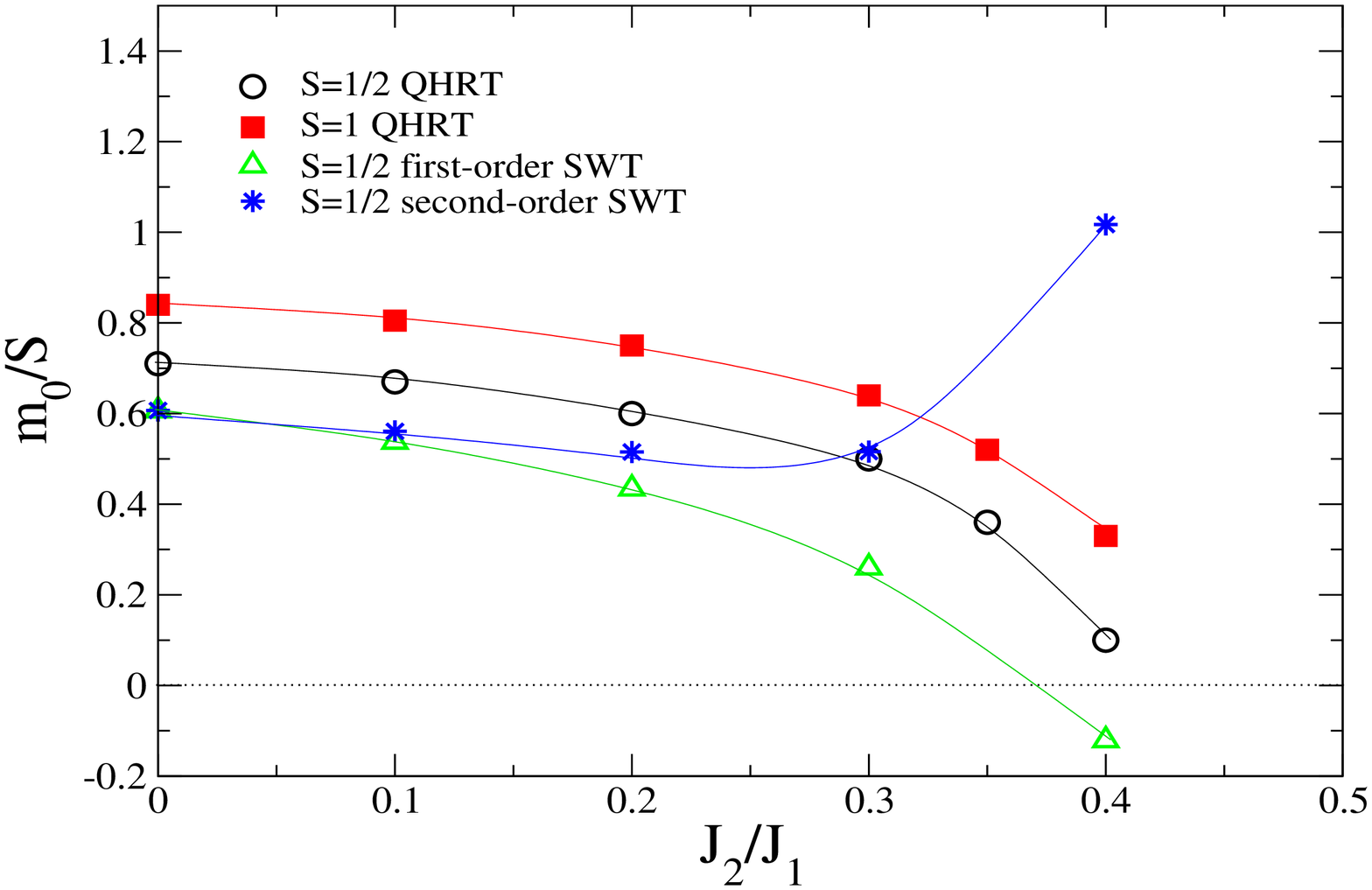,width=8.cm}
\caption{Extrapolated ground state staggered magnetization $m_0$ divided by its 
classical value $S$.  QHRT results for $S=1/2$ (empty circles) and $S=1$ (full squares). 
First-order (empty triangles) and second-order (stars) SWT results from \cite{igarachi}. 
Lines are a guide for the eyes.
\label{magnetiz}
}
\end{center}
\end{figure}

Our estimates of the ground state staggered magnetization, based on Fig. \ref{evolQ}, are 
reported in Table \ref{magtab} for both $S=1/2$ and $S=1$. The same data are shown in 
Fig. \ref{magnetiz}, together with the first and second order spin wave theory 
results \cite{igarachi}. As already discussed in Ref.\cite{ap}, at $\alpha=0$ and $S=1/2$
the QHRT estimate is very close to the commonly accepted value $m_0 \sim 0.307$ ~\cite{beard}. 
However, while first order SWT predicts a vanishing 
magnetization at $\alpha_c \sim 0.38$, our results indicate that the ground state magnetization 
extrapolates to zero only at significantly larger values of $\alpha \sim 0.43$.

At zero temperature, in a symmetry broken phase, the spin stiffness $\rho_s$ can be defined
in terms of the small momentum behavior of the transverse, static structure factor: 
\be
S_{\perp}(k)\to \frac{\rho_s}{2c_s} k  
\label{rhos0}
\ee
Within QHRT, the spin velocity $c_s$ has been previously 
evaluated from Eq. (\ref{dispersione}) while $S_{\perp}(k)$ is easily obtained from
the parametrization (\ref{correl}). As a result we obtain the simple analytical relationship:
$\rho_s= m_0^2 (J_1 - 2J_2)$ which, by use of Tab. \ref{magtab} provides an estimate of
the stiffness. The  proportionality between $\rho_s$ and $m_0^2$ guarantees that at the 
quantum critical point, where the order parameter $m_0$ vanishes, the spin stiffness 
necessarily tends to zero, as expected.

\section{TEMPERATURE DEPENDENCE OF PHYSICAL PROPERTIES}

\begin{figure}
\vspace{2mm}
\begin{center}
\epsfig{figure=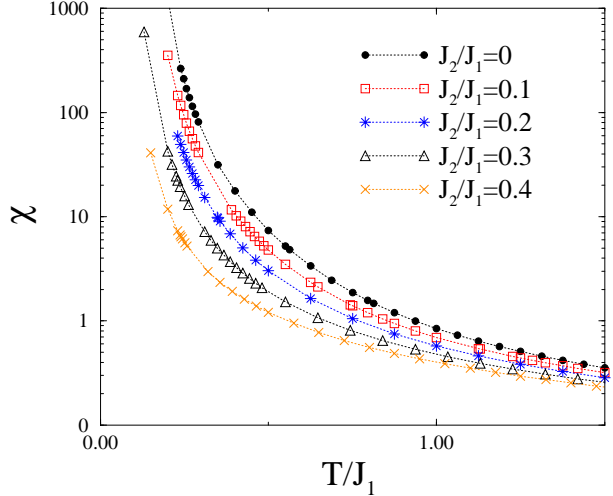,width=8cm}
\caption{$S=1/2$ low temperature staggered susceptibility versus temperature, 
for various values of the frustration \label{chi} at $m=0$ ($\chi^{s}_{\perp}=\chi^{s}_{||}$).
Lines are guides for the eyes.
}
\end{center}
\end{figure}

According to the definition of the spin-spin correlation function, we can define 
the staggered magnetic susceptibility in terms of the zero frequency value 
of $F^{\alpha\beta}({\bf k},\omega)$ at ${\bf k}=\bg$
\begin{eqnarray}
\chi^{s}_{\perp}&=&F^{xx}(\bg,0)=(\mu_{\perp}-4(J_1 - J_2))^{-1}\nm\\
 \chi^{s}_{||}&=&F^{zz}(\bg,0)=(\mu_{||}-4(J_1 - J_2))^{-1}.
\label{susk}
\end{eqnarray}
The results at $m=0$ are plotted in Fig. \ref{chi} as a function of temperature for
several values of the frustration ratio. 
In agreement with Mermin-Wagner theorem, true long-range order is absent at 
any finite temperature in two dimensions and in fact the susceptibility obtained via 
QHRT remains finite at every $T\ne 0$. Conversely, the divergence of $\chi$ when 
zero temperature is approached is consistent with the onset of long range antiferromagnetic 
order in the ground state of the model. 
In order to better analyze such a divergence we will refer to a closely related 
quantity: the antiferromagnetic correlation length, $\xi$ defined as $|\bk_{P}|^{-1}$, 
where $\bk=\bg+i\bk_{P}$ is the location of the poles of the function $F^{\alpha \beta}(\bk,\omega=0)$. 
The same $\xi$ essentially governs the exponential decay of the equal-time 
correlation function at $T \ne 0$. Close to the critical value of the wave vector 
$\bk=\bg$, the two-point correlation function at $\bk=\bg+\bq$ and $\omega=0$
takes the following form
\begin{equation}
F^{\alpha \beta}(\bg+\bq,0)\sim\frac{1}{\mu -4(J_1 - J_2) + q^2(J_1-2J_2)}
\end{equation}
leading to a correlation length algebraically related to the 
susceptibility (\ref{susk}) and simply expressed as a function of 
the parameter $\mu$:
\begin{equation}
\label{xi}
\xi^{-2}=\frac{\mu - 4(J_1 - J_2)}{J_1-2 J_2}
=\frac{1}{\chi^s(J_1-2 J_2)}
\end{equation}
The numerical values of $\xi(T)$ can be easily deduced from Fig. \ref{chi} and 
clearly show a strong suppression of long range correlations when the frustration is increased.
As predicted by weak-coupling (one-loop) RG analysis of the usual NLSM \cite{chakra}, 
when the temperature is lowered the system  
enters the classical renormalized regime characterized by an exponential 
divergence of $\xi$. In this region of the phase diagram
the correlation length diverges exponentially, just as it would in the classical model: 
\begin{equation}
\label{expo}
\xi \sim \exp(2 \pi \tilde\rho_s /T)
\end{equation} 
where the coefficient $\tilde\rho_s$ is identified through the weak coupling RG analysis, with
the spin-stiffness renormalized by quantum fluctuations\cite{ivanov,chubukov,sachdev}. 
\begin{figure}
\vspace{2mm}
\begin{center}
\epsfig{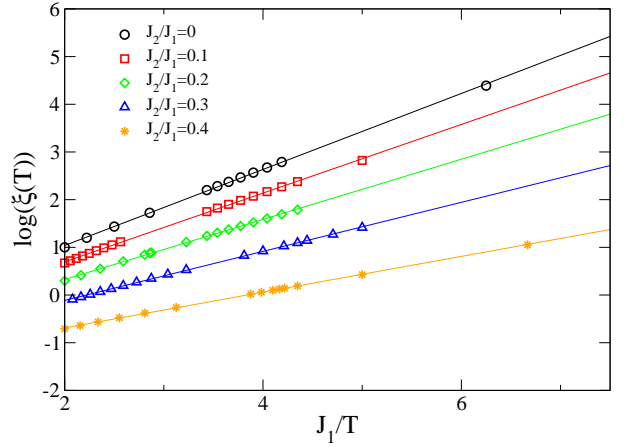}
\caption{\label{logcsi}
$\log(\xi(T))$ versus the inverse temperature $J_1/T$ for the $S=1/2$ system and 
different values of the frustration ratio. Straight lines are linear fits of the 
low temperature data, which clearly follow the exponential law (\ref{expo}). Qualitatively 
similar results have been obtained for $S=1$.}
\end{center}
\end{figure}
As shown in Fig. (\ref{logcsi}), the QHRT equation well reproduces, for different values 
of the frustration, the exponential behavior of $\xi$ at low temperature as already
found in Ref. \cite{ap}. 
Eq. (\ref{expo}) is valid in the low temperature limit ($T<<J$) when correlations at 
the antiferromagnetic wave vector become long ranged. In this regime 
it is possible to estimate the renormalized stiffness coefficient $\tilde\rho_s$ 
which governs the exponential divergence of the correlation length (\ref{expo}) 
via a linear fit of the data in Fig. \ref{logcsi}.

%
\begin{figure}
\vspace{2mm}
\begin{center}
\epsfig{figure=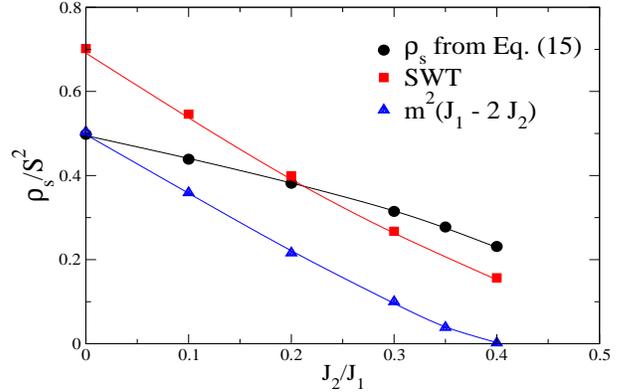,width=8cm}
\caption{$S=1/2$ renormalized spin stiffness as a function of the frustration. 
We report the values obtained from Eq. (\ref{rhos0}) (triangles), the coefficient $\tilde\rho_s$ from the low temperature divergence  of the correlation length, Eq. (\ref{expo}) (circles) and the SWT estimate (squares) from ~\cite{ivanov}. 
}
\label{stiff}
\end{center}
\end{figure}
The results of such a fit are shown in Fig. \ref{stiff}, together with the direct evaluation of the stiffness $\rho_s$ via Eq. (\ref{rhos0}) at $S=1/2$. 
Frustration depresses the correlation length, nevertheless $\tilde\rho_s$ 
does not vanish in the range of frustration $[0,0.4]$ we
have investigated and no clear sign of quantum critical point can be inferred from these
data. A similar result was actually found in Ref. \cite{ivanov} using variational 
spin-wave approach while exact diagonalization studies, followed by finite size 
scaling,  \cite{feiguin,poil} are not conclusive, the spin stiffness estimate depending on the
chosen cluster geometry. 

The comparison between $\tilde\rho_s$ and the direct evaluation of the stiffness $\rho_s$ via 
Eq. (\ref{rhos0}) (see  Fig. \ref{stiff}) clearly shows that, within
QHRT, the exponential divergence of the correlation length at low temperature is not
governed by the same quantity which appears in the zero temperature transverse structure factor,
contrary to the prediction of the RG analysis of the non linear sigma model \cite{zinn}. 
In order to investigate the origin of such a discrepancy, we performed a careful 
analysis of the QHRT equation in the classical limit
at $J_2=0$, which corresponds to the $S\to\infty$ case. At zero temperature, the ground state
magnetization is clearly $m_0=S$, providing the correct limiting result for the 
classical spin stiffness: $\rho_s=S^2J_1$. Instead, the numerical
integration of the evolution equation gives a diverging correlation length consistent 
with Eq. (\ref{expo}) where the renormalized stiffness is given by $\tilde\rho_s=S^2J_1/3$. 
This estimate actually coincides with the lowest order term of a large $N$ expansion in 
the SU$(N)$ non linear sigma model \cite{sachdev} evaluated at $N=3$. The correct RG result 
for the SU$(N)$ non linear sigma model, valid at all orders in $N$, is instead given by 
$\tilde\rho_s=S^2J_1/(N-2)$ which, at $N=3$ coincides with the zero temperature evaluation of $\rho_s$. 
A careful analysis of the RG approach in $2+\epsilon$ dimension 
\cite{zinn} suggests that the origin of the discrepancy found in the QHRT equations may be traced 
back to the chosen parametrization of the correlation functions (\ref{correl}). The reason is twofold:
\par\noindent
$\bullet$ In the QHRT equation (\ref{evoluzione}) both longitudinal and transverse 
correlations contribute
to one loop order, while in the RG approach only the transverse propagator enters the beta-function.
Due to the non linear sigma model constraint, at low temperature the longitudinal correlations can be
expressed in terms of the transverse ones:
\begin{eqnarray}
F^{zz}(r) &\sim&\,\Big [ <[1-\pi(r)^2]^{\frac{1}{2}}[1-\pi(0)^2]^{\frac{1}{2}}> - \nonumber \\
&-& <[1-\pi(0)^2]^{\frac{1}{2}}>^2 \Big ] \nonumber \\
&\sim& \, \frac{1}{4}\left [<\pi(r)^2\pi(0)^2> -<\pi(0)^2>^2\right ] \nonumber \\
&\sim& \frac{1}{2}\left [F^{xx}(r)\right ]^2
\end{eqnarray}
where $\pi(r)$ identifies transverse spin fluctuations. 
\par\noindent
$\bullet$ Moreover, within our approximation (\ref{correl}) the wave-vector dependence of the correlation
function is not renormalized by fluctuations, while the RG analysis in $2+\epsilon$ dimensions
shows that the one loop corrections do affect the momentum dependence of the one particle 
irreducible two point function \cite{zinn}:
\begin{equation}
\Gamma(p)= \frac{\Lambda^\epsilon}{T}\left [p^2
\left (1+\frac{T}{2\pi\epsilon}\right) +
h\left (1+\frac{(N-1)T}{4\pi\epsilon}\right) \right ]
\label{pert}
\end{equation}
where $\Lambda$ is the ultraviolet cut-off of the $O(N)$ non linear sigma model. 
This expression should be compared with the long wavelength limit of the 
corresponding QHRT approximation:
\begin{equation}
\Gamma(\bp)=\beta\mu_\perp + \beta\tilde w(\bp+\bg)
\end{equation}
showing that in QHRT the momentum dependence is fixed by the spin-spin interaction.

These remarks suggest that a more flexible parametrization
of the spin correlations should be used in the QHRT equations in order to achieve a quantitative
representation of the low temperature divergence of the correlation length. 

In order to better understand the effects of frustration on 
the onset of N\'eel order, we also investigated the $\alpha$ dependence of the 
temperature $T_\times$ below which Eq. (\ref{expo}) is valid. In the following we refer to 
$T_\times$ as crossover temperature. 
At mean-field level we known that the inclusion of a n.n.n. interaction increases 
the bare value of the coupling constant $g$ and tends to destroy the N\'eel order. When the 
system approaches the quantum critical point, controlled by the $g=g_c$ fixed point, 
the RG picture (qualitatively) predicts a decrease of the crossover temperature.
\begin{figure}
\vspace{2mm}
\begin{center}
\epsfig{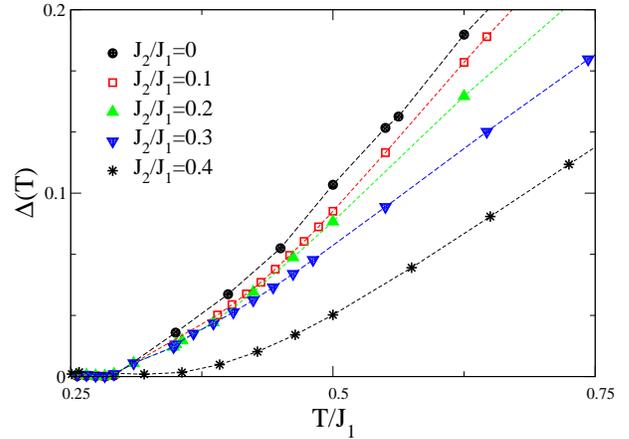}
\caption{\label{cross}
Plot of the function $\Delta(T)$ defined as $\Delta(T)=|\log(\xi(T)) - l_{fit}(T)|$, where 
$l_{fit}(T)$ is the linear fit function describing the asymptotic behavior of $\log\xi(T)$ 
at low $T$. The function is plotted for different value of the frustration ratio. 
}
\end{center} 
\end{figure}
The crossover temperature can be estimated by studying the temperature dependence of the 
function  $\Delta(T)$  defined as $\Delta(T)=|\log(\xi(T)) - l_{fit}(T)|$, where $l_{fit}(T)$ 
is the linear fit function describing the asymptotic behavior of $\log\xi(T)$ at low $T$: 
$\Delta(T)$ then measures the deviations of $\xi(T)$ from the exponential law (\ref{expo}).
Figure \ref{cross} clearly shows that it is possible to identify a crossover temperature 
$T_\times$ by the requirement $\Delta(T)\approx 0$ for $T<T_\times$.
The results reported in Fig. \ref{cross} seem to indicate that $T_\times$  does not  
depend much on $\alpha$, its value remaining basically unchanged when frustration is increased.  
The weak dependence of the crossover temperature on frustration may suggest
either a very narrow critical region near the quantum transition, or a shift in the quantum critical 
point toward larger values of $\alpha > 0.5$ \cite{ivanov}. 
\begin{figure}
\vspace{2mm}
\begin{center}
\epsfig{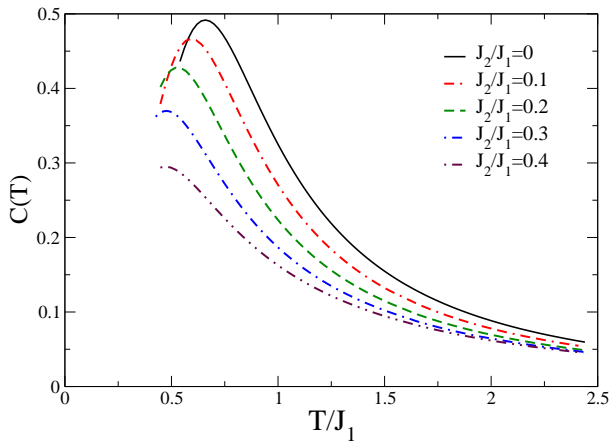}
\caption{\label{calore}
Zero field specific heat for $S=1/2$ }
\end{center} 
\end{figure}

Finally, from the free energy we calculated zero field specific heat $C(T)$ for $S=1/2$ and different 
frustrations, shown in  Fig. (\ref{calore}). As expected, frustration play a minor role at 
high temperature, the specific heat showing a paramagnetic behavior as $1/T^2$ for all values 
of the coupling J$_2$. The position of the maximum  moves at low temperatures and its intensity 
decreases for increasing value of the frustration, in agreement with  SWT ~\cite{ivanov} 
and small-size Exact Diagonalization results ~\cite{dagotto}.

\section{CONCLUSIONS}

We have studied finite temperature properties of the frustrated Heisenberg model on 
a square lattice in the N\'eel phase,  by use of the quantum hierarchical reference 
theory, a non-perturbative implementation of the renormalization group method.
We derived an exact evolution equation which describes how the free energy of the system 
changes when antiferromagnetic fluctuations over growing wavelengths are included. 
This equation is supplemented by an approximate representation of the dynamical correlation
functions of the model (\ref{correl}). 
Within the limitations induced by our parameterization of the two-point correlations,
the method employed does not suffer from uncertainties in the 
extrapolation to the thermodynamic limit \cite{comment}. Our results are valid 
in a wide range of temperature and for all spin values because our approach does not rely 
on any mapping on effective long wavelength theory and preserves the microscopic features 
of the model Hamiltonian. We focused on the role of fluctuations of wavevector 
close to $\bg=(\pi,\pi)$ and then we could not investigate the possible occurrence of other order 
parameters characterized by different periodicities, as in spiral or collinear phases. 
 
Starting from the knowledge of the free energy of the system, we calculated spin 
susceptibility, two point correlation function  and correlation length at finite temperature 
and we also estimated ground state properties such as spin stiffness and 
staggered magnetization by extrapolation to the $T\to 0$ limit. Calculations have been performed 
for spin $S=1/2$ and $S=1$ and different values of the frustration ratio, in order 
to investigate the role and extent of quantum corrections on top of the semiclassical spin-wave picture.
The investigated range of frustration is restricted to $\alpha < 0.4$ because
beyond this point, fluctuations at wavevector $(\pi,0)$ are expected to become important and
our simple parametrization of the dynamical correlation functions cannot be retained. 

The spontaneous staggered magnetization extrapolates to zero at about $\alpha\sim 0.43$ 
for $S=1/2$ while a stronger frustration is required for $S=1$. Within the QHRT approach, 
both the spin velocity and stiffness are related to the order parameter and therefore are
expected to vanish at the quantum critical point. For comparison we recall that 
in the non linear sigma model representation of the quantum antiferromagnets, 
when the $T=0$ critical point is approached, both the staggered magnetization 
and the spin stiffness vanish, while the spin velocity is expected to remain finite. 
Our results indicate that frustration strongly suppresses the correlations at finite 
temperature  but, at the same time, the value of the crossover temperature $T_\times$ remains 
basically unchanged when the frustration is increased. 
This fact may suggest a very narrow critical region near the quantum transition 
or a shift of the quantum critical point toward larger values of $\alpha > 0.5$. 
The estimate of the renormalized stiffness $\tilde\rho_s$ obtained via the low temperature divergence of 
the correlation length, Eq. (\ref{expo}), does not vanish in the whole frustration range we investigated
and is remarkably larger than the direct evaluation of the stiffness $\rho_s$ 
obtained through the zero temperature static 
structure factor (\ref{rhos0}). The equality of these two quantities 
is commonly expected on the basis of the known behavior of the SU$(N)$ sigma-model 
in the $N\to \infty$ limit and the RG analysis in $2+\epsilon$ dimension \cite{zinn}.
The difference in the two estimates of $\rho_s$ 
we found in the SU$(2)$ case is most likely a consequence of 
the approximate closure (\ref{correl}) we have studied, as shown by a detailed analysis of the 
classical limit of the QHRT equations. In order to go beyond the approximation
examined in this work, a representation of the low temperature spin correlations 
richer than those commonly adopted in the many body literature \cite{varie} is needed. 

Nowadays, real materials displaying the properties of the $J_1-J_2$ model in the N\'eel phase
have not been yet synthesized. Nevertheless, the recent experiments at high pressure 
performed on the Li$_2$VOSiO$_4$ \cite{pietro3} seem to indicate that a wide range of 
the $J_1-J_2$ phase diagram can be studied  by applying high hydrostatic pressure, which modifies 
the frustration ratio $J_2/J_1$ by changing the bond lengths and angles.
Although the relationship between pressure and frustration ratio is not obvious at all,
we believe that the results of our investigation may be useful for the interpretation of 
future neutron scattering or NMR-NQR experiments. In addition ground state configuration of the  VOMoO$_4$ compound is still a controversial issue, first principle calculation of the J$_2$ coupling being in disagreement with the estimates based on the NMR-NQR experiments. Specific heat data ~\cite{pietro3} for the VOMoO$_4$ compound may be analyzed on the basis of our finite-temperature results,  the position of the maximum in the zero-field specific heat giving an alternative way for estimating the frustration parameter.

We thank L. Reatto and P. Carretta for useful discussions. The financial 
support of MIUR through PRIN-2003 is also acknowledged.


\begin{thebibliography}{99}
\bibitem{note} Here, for simplicity, we consider spatially isotropic systems. The correspondig
expressions appropriate for a lattice model will be given later.
\bibitem{comment} As pointed out in \cite{eina}, for a finite system even a disordered
(spin liquid) phase may have a finite stiffness.
\end{thebibliography}
\end{document}